\documentclass[journal]{IEEEtran}

\usepackage{amsmath,amsfonts}
\usepackage{algorithmic}
\usepackage{algorithm}
\usepackage{array}
\usepackage[caption=false,font=normalsize,labelfont=sf,textfont=sf]{subfig}
\usepackage{textcomp}
\usepackage{stfloats}
\usepackage{url}
\usepackage{verbatim}
\usepackage{graphicx}
\usepackage{cite}
\usepackage{color}
\usepackage{booktabs}
\ifCLASSINFOpdf
\else
\fi
\begin{document}
%
\title{Multi-mode OAM Convergent Transmission with Co-divergent Angle Tailored by Airy Wavefront}
%
%
%

\author{Yufei ZHAO,
        Ziyang WANG,
        Yilong LU,~\IEEEmembership{Fellow,~IEEE,}
        and~Yong Liang GUAN,~\IEEEmembership{Senior Member,~IEEE}
\thanks{This research is supported by the National Research Foundation, Singapore and Infocomm Media Development Authority under its Future Communications Research \& Development Programme Grant No. FCP-NTU-RG-2021-015. It was also supported by A*STAR under its RIE2020 Advanced Manufacturing and Engineering (AME) Industry Alignment Fund - Pre Positioning (IAF-PP) Grant No. A19D6a0053. Any opinions expressed in this material are those of the authors and do not reflect the views of the funders. (Corresponding author: Yong Liang Guan, e-mail: eylguan@ntu.edu.sg).}
\thanks{Yufei ZHAO, Yilong LU and Yong Liang GUAN are with the School of Electrical and Electronic Engineering, Nanyang Technological University, 639798, Singapore.

Ziyang WANG is with Research Institute for Frontier Science, Beihang University, Beijing 100191, China.}
\thanks{Manuscript received August 26, 2022; major revised November 10, 2022; Accepted March 6, 2023}}

\maketitle

\begin{abstract}
Wireless backhaul offers a more cost-effective, time-efficient, and reconfigurable solution than wired backhaul to connect the edge-computing cells to the core network. As the amount of transmitted data increases, the low-rank characteristic of Line-of-Sight (LoS) channel severely limits the growth of channel capacity in the point-to-point backhaul transmission scenario. Orbital Angular Momentum (OAM), also known as vortex beam, is considered a potentially effective solution for high-capacity LoS wireless transmission. However, due to the shortcomings of its energy divergence and the specificity of multi-mode divergence angles, OAM beams have been difficult to apply in practical communication systems for a long time. In this work, a novel multi-mode convergent transmission with co-scale reception scheme is proposed. OAM beams of different modes can be transmitted with the same beam divergent angle, while the wavefronts are tailored by the ring-shaped Airy compensation lens during propagation, so that the energy will converge to the same spatial area for receiving. Based on this scheme, not only is the Signal-to-Noise Ratio (SNR) greatly improved, but it is also possible to simultaneously receive and demodulate OAM channels multiplexed with different modes in a limited space area. Through prototype experiments, we demonstrated that 3 kinds of OAM modes are tunable, and different channels can be separated simultaneously with receiving power increasing. The measurement isolations between channels are over 11 dB, which ensures a reliable 16-QAM multiplexing wireless transmission demo system. This work may explore the potential applications of OAM-based multi-mode convergent transmission in LoS wireless communications.
\end{abstract}


%
\IEEEpeerreviewmaketitle

\section{Introduction}
%
%
%
%
\IEEEPARstart{T}{o} support data-rich and delay-sensitive applications such as connected autonomous vehicles, smart mobility, smart ports, and extended reality, the 5th generation (5G) and 6th generation (6G) wireless networks are deemed to have a massive deployment of edge-computing small cells. Wireless backhaul offers a more cost-effective, time-efficient, and reconfigurable solution than wired backhaul to connect the small cells to the core network \cite{MIMO-backhaul}. However, conventional wireless backhaul links face practical issues such as limited spectrum and installation space on the radio tower, and the space division multiplexing technology cannot effectively address the contradiction between capacity and resource in a point-to-point Line-of-Sight (LoS) scenario \cite{MIMO-backhaul}. As an inherent property of Electro-Magnetic (EM) waves, Orbital Angular Momentum (OAM) beams have attracted a lot of attention \cite{chao1,lee,tie}. The low correlation between OAM modes brings additional spectrum efficiency and energy efficiency \cite{Allen}. In the last decade, OAM has been employed in the Radio Frequency (RF) and expected to be utilized in the high capacity LoS transmission radiated by the specific traveling-wave antennas \cite{xiong} or antenna array \cite{chen2}.

\begin{figure}[hbt]
\centering
\includegraphics[width=3.5in]{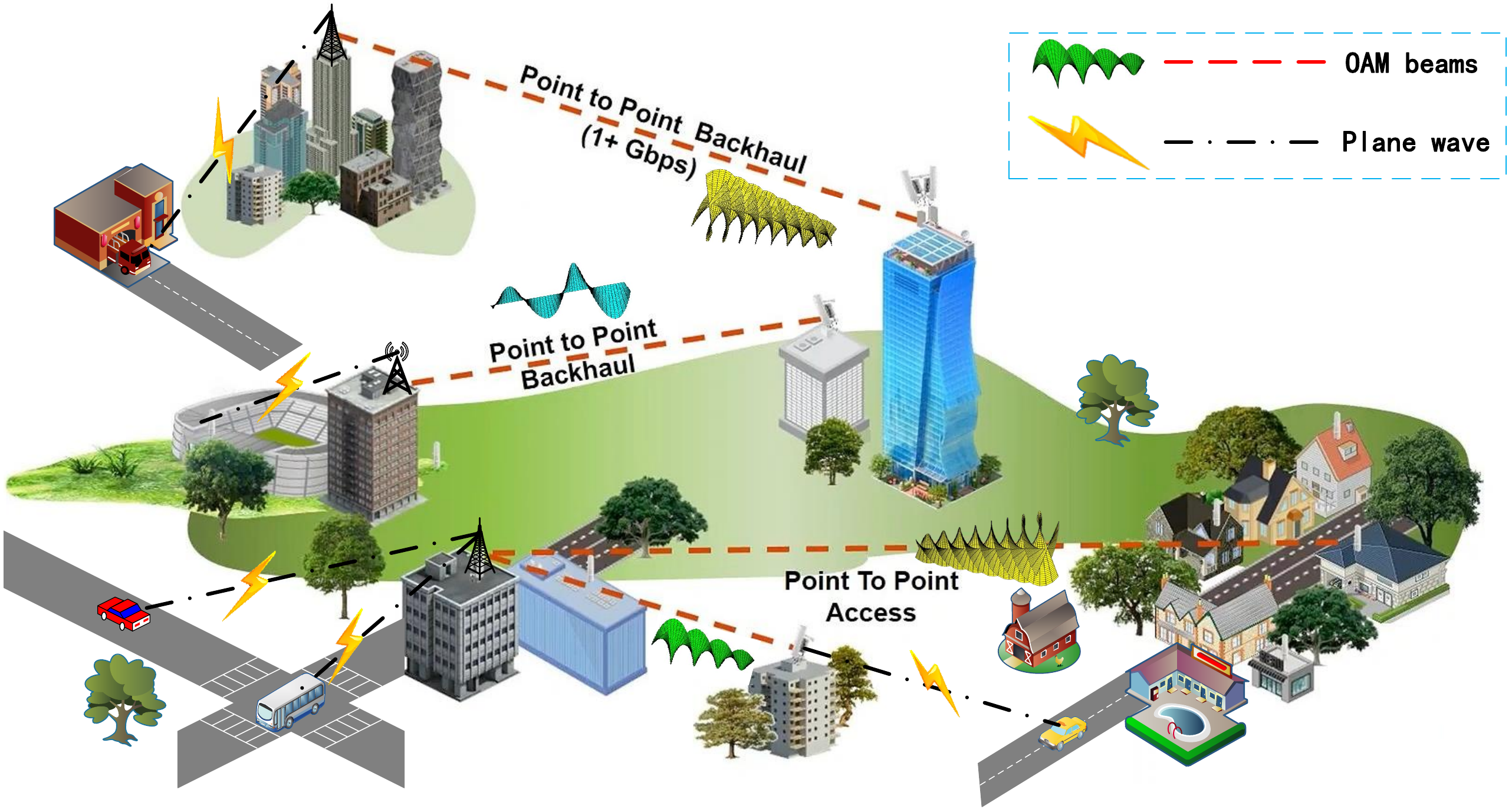}
\caption{LoS scenarios for OAM wireless transmission.}
\label{fig0}
\end{figure}

As wireless communications gradually expand into higher frequency bands, e.g., mmWave, terahertz, scenarios for LoS transmission become more common. As shown in Fig. \ref{fig0}, in the primarily LoS scenario, such as the wireless backhaul links, the multiplexing channel is highly correlated due to the limited scattering in the propagation environment, resulting in a rank-deficient channel matrix \cite{MIMO-backhaul}. Hence, OAM beams, with their rich and multifaceted wavefront characteristics, introduce a new variant to the monotonous LoS communication scenario \cite{xiong}. However, the various drawbacks of OAM beams generation and propagation have also frustrated researchers.

As we know, a typical OAM beam propagates in an inverted cone shape with its main lobe energy distributed over a circular ring \cite{yufei1}. As the transmission distance increases, the radius of the ring expands, and the power received in a limited space gradually decreases. Actually, due to its inherent spiral wavefront phase distribution characteristics, there is always a central energy hole in the OAM beam, which considerably limits the energy efficiency of a Tx-Rx system, especially for long-distance transmission. Moreover, as the mode order increases, the beam divergence problem becomes more serious, which is why most OAM beam-based transmission experiments generally use low-order modes \cite{Liang}. However, the pace of research exploration has not been slowed down by these challenges. In recent years, some researchers have proposed the concept of planar spiral OAM. The main lope of the radiation pattern is along the transverse plane, which means that it can avoid the impact of the central null to some extent \cite{PSOAM}. However, due to the cumbersome resonant cavity excitation structure, it is still difficult to realize more modes for multiplexing transmission, which means that it is still challenging to apply this method to practical communication or detection systems.

In \cite{malu1}, C. Zhang $\textit{et. al.}$ propose that the rotation angle Doppler characteristics can be used to detect OAM modes after long-distance transmission. Based on this single-point detection method, the OAM multi-mode index mapping long-distance transmission technology has also been proposed and verified by experiments \cite{yufei-twc}. In 2021, Y. Zhang $\textit{et. al.}$ propose a novel partial arc sampling receiving scheme based on antenna array\cite{Yanming}. However, these receiving methods do not inherently solve the problem of OAM beam energy divergence at the transmitter. Signal processing algorithms at the receiver still suffer from low SNR. To deal with this problem, S. Gao $\textit{et. al.}$ design a kind of bifocal lens to converge OAM beams, which has a good effect on reducing the beam divergence angle. Moreover, Y. Zhao $\textit{et. al.}$ propose a scheme to converge and separate multi-mode OAM beams longitudinally along the propagation \cite{yufei2}. In 2021, based on an Airy-OAM-phase modulation method, Y. Huang $\textit{et. al.}$ design a single-layer reflectarray that can transform the plane wave to a vortex beam and auto-focus it to a certain focal point \cite{xiuping}.

Since most of the works only focus on the design of the RF front-end, ignoring the fundamental requirements of the multi-mode multiplexing transmission of the communication system itself, there are still problems to be solved urgently. For example, the value of OAM beams for communication lies in the multi-mode flexible modulation and multiplexing transmission capabilities \cite{NTT}. If the generation of vortex wavefront and the convergence phase steering are performed on the same dielectric lens or metasurface, we cannot use different OAM modes to transmit different information independently, which will greatly reduce the application value of OAM beams, especially for wireless communications.

In this paper, we propose a novel scheme for multi-mode OAM beams convergence transmission with co-divergence angles. The formation of vortex wavefronts and the steering of convergence beams are two separate processes. Each OAM mode is connected with one baseband modulation channel independently. At the RF transmitter end, multi-mode OAM beams can be generated by various methods, e.g., \cite{Liang,weite}. By adjusting the radius of the transmitting antennas, OAM beams of different modes can share the same beam divergence angle and propagation path. Then, the OAM beams are tailored by the circular Airy wavefront, and converged to the same focal point along the propagation axis. As a result, the receiver only needs to be placed at the focal position to receive signals carried by different OAM beams simultaneously. Due to the obvious non-diffraction and self-focusing properties (finite-size generators only obtain limited quasi-nondiffracting propagation depths) of circular Airy beams \cite{Berry,R4}, the originally divergent OAM beams suddenly converge to the same spatial position during the propagation process, thereby greatly improving the SNR of receiving signals. Both simulations and practical communication experiments are used in this paper to validate the correctness and effectiveness of this scheme, which has reference significance for future LoS wireless communication scenarios.

The rest of this paper is organized as the following. Section \ref{sect2} shows the details of the OAM generator with co-divergence angles. Section \ref{sect3} gives the architecture and mathematical model of the circular Airy wavefront tailored convergent transmission scheme. Section \ref{sect4} describes the experimental prototype in a microwave anechoic chamber, and makes the performance evaluation by simulation and measurement results. Then, the conclusion is drawn in Sect. \ref{sect5}.


%

\section{OAM Generator with Co-divergent Angles} \label{sect2}
As we know, generally, OAM beams of different modes have inconsistent beam divergence angles. In other words, the main-lobe directions of different OAM beams differ from each other, which makes it difficult for the receiving end to simultaneously receive the maximum energy of all OAM beams at the same position. In \cite{R7}, the authors hope to illuminate the target with multiple different OAM beams to obtain radar echoes of different OAM modes, which requires that the multi-mode OAM beams have the same main-lobe radiation direction. To deal with that, authors from \cite{R7} propose using a concentric Uniform Circular Array (UCA) system to generate multiple OAM beams. The main-lobe direction of various OAM beams can be adjusted by changing the array radius. Similarly, the authors of \cite{R8} also apply the concentric UCA method to OAM-based radar imaging scenarios, which helps them to collimate the beams of different OAM modes to the same elevation direction to increase the echo power. For simplicity without losing generality, we also take the concentric UCA as our multi-mode OAM beams generator in this paper.

\begin{figure}[hbt]
\centering
\includegraphics[width=3.2in]{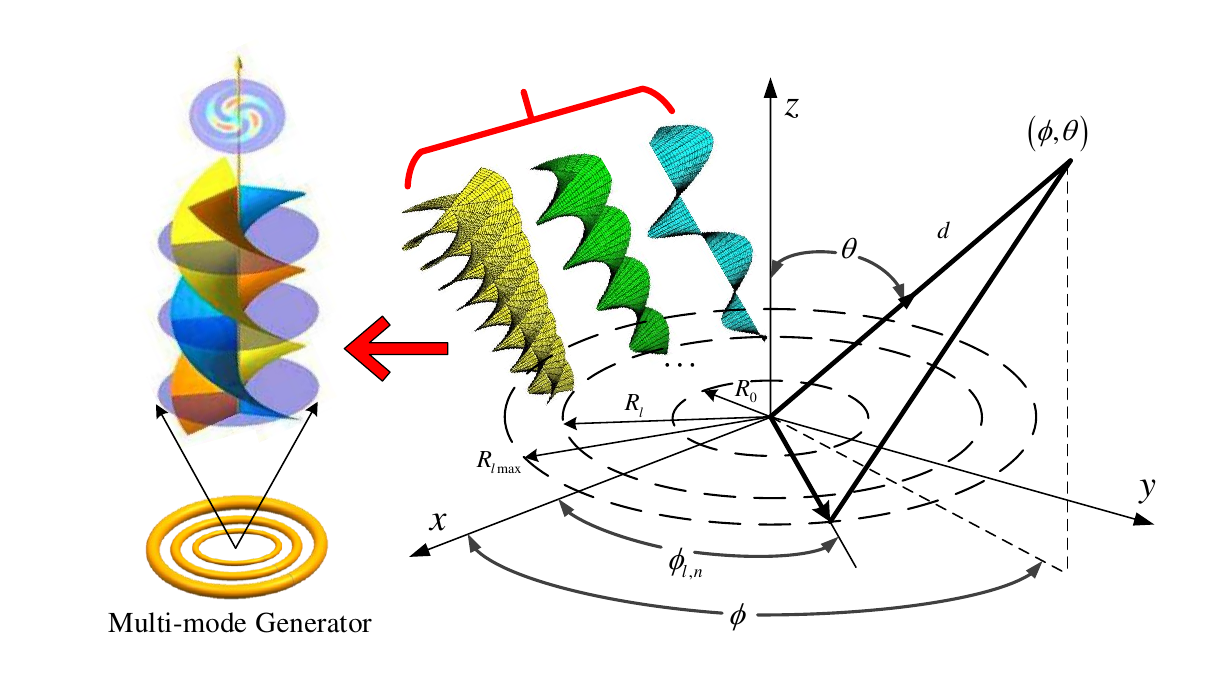}
\caption{OAM beams co-divergent angle transmission.}
\label{fig1}
\end{figure}

Different from \cite{R7}, \cite{R8}, the concentric UCA is only used as the signal transmitter in our communication system. In this communication scenario, the OAM beams do not need to be reflected by the target, but are directly sampled and received by the receiver after propagation. Additionally, to confirm the concentric UCA system's band-pass modulation data transmission capabilities, we also design and validate a principle prototype through both simulation and practical experimentation.
As shown in Fig. \ref{fig1}, assuming that there are $N$ isotropic elements spaced on a ring equally with radius ${R_l}$, the radiation pattern of each UCA can be denoted as
\begin{equation} \label{eq1}
{{\vec E}_l}\left( {d,\theta ,\phi } \right) = \frac{{{e^{ - jkd}}}}{d}\sum\limits_{n = 0}^{N - 1} {{{\vec A}_{l,n}}{e^{jk{R_l}\sin \theta \cos \left( {{\phi _{l,n}} - \phi } \right)}}} ,
\end{equation}
where, $k = {{2\pi } \mathord{\left/ {\vphantom {{2\pi } \lambda }} \right. \kern-\nulldelimiterspace} \lambda }$ is the wave vector, $\lambda $ is the wavelength, $l$ is an integer denoting the OAM mode, $\left\lfloor { - N/2} \right\rfloor  \le l < N/2$ \cite{yufei-EL}, $j$ is the imaginary unit. In the cylindrical coordinate system, $\theta $ and $\phi $ denotes the pitch and azimuth independent variables of the radiation pattern, ${\phi_{l,n}} = 2\pi \left( {{n \mathord{\left/{\vphantom {n N}} \right.\kern-\nulldelimiterspace} N}} \right)$ is the azimuth angle of the $n$-th element on the UCA ring, ${{\vec A}_{l,n}} = {I_{l,n}}{e^{j{\alpha _{l,n}}}}$ is the signal incentive coefficient of the $n$-th array element of the $l$-mode, ${I_{l,n}}$ and ${\alpha _{l,n}}$ denote signal amplitude and phase respectively.

In order to generate the $l$-mode OAM beam, the phase excitation ${\alpha _{l,n}}$ of the $n$-th element needs to be satisfied as
\begin{equation}
{\alpha _{l,n}} = l \cdot \left[ {2\pi \left( {{n \mathord{\left/
 {\vphantom {n N}} \right.
 \kern-\nulldelimiterspace} N}} \right) + {\varphi _l}} \right],
\end{equation}
where, ${{\varphi _l}}$ denotes the phase adjusting factor for the $l$-mode, which is the same for each element of one OAM mode, similarly, ${I_{l,n}}$ is equal for each array element. For other kind of OAM generators, ${{\vec A}_{l,n}}$ also denotes the feeding signal for each OAM mode. For clarify, let's assume that ${{\varphi _l}}=0$, ${I_{l,n}}=1$. Assuming that there are many elements on the UCA ring, the summation of (\ref{eq1}) can be approximated as an integral over $2\pi$ angle range, which is derived as follows,
\begin{equation} \label{eq2}
\begin{array}{*{20}{l}}
  {{E_l}\left( {d,\theta ,\phi } \right) = \frac{{{e^{ - jkd}}{e^{jl\phi }}}}{d}\sum\limits_{n = 0}^{N - 1} {{e^{jk{R_l}\sin \theta \cos \left( {{\phi _{l,n}} - \phi } \right)}}{e^{jl\left( {{\phi _{l,n}} - \phi } \right)}}} } \\
  \begin{gathered}
  {\kern 1pt} {\kern 1pt} {\kern 1pt} {\kern 1pt} {\kern 1pt} {\kern 1pt} {\kern 1pt} {\kern 1pt} {\kern 1pt} {\kern 1pt} {\kern 1pt} {\kern 1pt} {\kern 1pt} {\kern 1pt} {\kern 1pt} {\kern 1pt} {\kern 1pt} {\kern 1pt} {\kern 1pt} {\kern 1pt} {\kern 1pt} {\kern 1pt} {\kern 1pt} {\kern 1pt} {\kern 1pt} {\kern 1pt} {\kern 1pt} {\kern 1pt} {\kern 1pt} {\kern 1pt} {\kern 1pt} {\kern 1pt} {\kern 1pt} {\kern 1pt} {\kern 1pt} {\kern 1pt} {\kern 1pt} {\kern 1pt} {\kern 1pt} {\kern 1pt} {\kern 1pt} {\kern 1pt} {\kern 1pt}  = \frac{{{e^{ - jkd}}{e^{jl\phi }}}}{d}\frac{N}{{2\pi }}\int_0^{2\pi } {{e^{jk{R_l}\sin \theta \cos \left( {{\phi _{l,n}} - \phi } \right)}} \cdots }  \hfill \\
  {\kern 1pt} {\kern 1pt} {\kern 1pt} {\kern 1pt} {\kern 1pt} {\kern 1pt} {\kern 1pt} {\kern 1pt} {\kern 1pt} {\kern 1pt} {\kern 1pt} {\kern 1pt} {\kern 1pt} {\kern 1pt} {\kern 1pt} {\kern 1pt} {\kern 1pt} {\kern 1pt} {\kern 1pt} {\kern 1pt} {\kern 1pt} {\kern 1pt} {\kern 1pt} {\kern 1pt} {\kern 1pt} {\kern 1pt} {\kern 1pt} {\kern 1pt} {\kern 1pt} {\kern 1pt} {\kern 1pt} {\kern 1pt} {\kern 1pt} {\kern 1pt} {\kern 1pt} {\kern 1pt} {\kern 1pt} {\kern 1pt} {\kern 1pt} {\kern 1pt} {\kern 1pt} {\kern 1pt} {\kern 1pt} {\kern 1pt}  \times {e^{jl\left( {{\phi _{l,n}} - \phi } \right)}}d({\phi _{l.n}} - \phi ) \hfill \\
\end{gathered}
\end{array}.
\end{equation}
Since the Bessel function of the first kind has the following form
\begin{equation}
{J_l}\left( z \right) = \frac{1}{{2\pi {j^l}}}\int_0^{2\pi } {{e^{jz\cos \vartheta }}{e^{jl\vartheta }}d\vartheta },
\end{equation}
hence, the integral in (\ref{eq2}) can be converted into the form of a Bessel function as
\begin{equation} \label{eq3}
\begin{array}{*{20}{l}}
  {{{\vec E}_l}\left( {d,\theta ,\phi } \right) = N{j^l}{d^{ - 1}}{e^{ - jkd}}{e^{jl\phi }}{J_l}\left( {k{R_l}\sin \theta } \right)} \\
  {{\kern 1pt} {\kern 1pt} {\kern 1pt} {\kern 1pt} {\kern 1pt} {\kern 1pt} {\kern 1pt} {\kern 1pt} {\kern 1pt} {\kern 1pt} {\kern 1pt} {\kern 1pt} {\kern 1pt} {\kern 1pt} {\kern 1pt} {\kern 1pt} {\kern 1pt} {\kern 1pt} {\kern 1pt} {\kern 1pt} {\kern 1pt} {\kern 1pt} {\kern 1pt} {\kern 1pt} {\kern 1pt} {\kern 1pt} {\kern 1pt} {\kern 1pt} {\kern 1pt} {\kern 1pt} {\kern 1pt} {\kern 1pt} {\kern 1pt} {\kern 1pt} {\kern 1pt} {\kern 1pt} {\kern 1pt} {\kern 1pt} {\kern 1pt} {\kern 1pt} {\kern 1pt} {\kern 1pt} {\kern 1pt} {\kern 1pt} {\kern 1pt}  = \gamma {J_l}\left( {k{R_l}\sin \theta } \right){e^{jl\phi }}}
\end{array}.
\end{equation}
Due to the limited size of the receiver, multi-mode OAM multiplexing beams need to be distributed in nearly the same spatial range. In order to have the same beam divergence angle ${{\theta}}$ for different OAM modes, the radius of the $l$-mode OAM generator (e.g., the UCA ring, loop waveguide, etc.) should be adjusted as
\begin{equation} \label{eq4}
{R_l} = {{{\chi _l}} \mathord{\left/
 {\vphantom {{{\chi _l}} {\left( {k\sin \theta } \right)}}} \right.
 \kern-\nulldelimiterspace} {\left( {k\sin \theta } \right)}},
\end{equation}
where ${{\chi _l}}$ denotes the abscissa corresponding to the maximum value of the ${l}$-order Bessel function \cite{liukang}. In this way, the main lobes of different OAM beams can be superimposed on the same annular phase plane by adjusting the radii of generators.

\begin{figure}[hbt]
\centering
\includegraphics[width=3.4in]{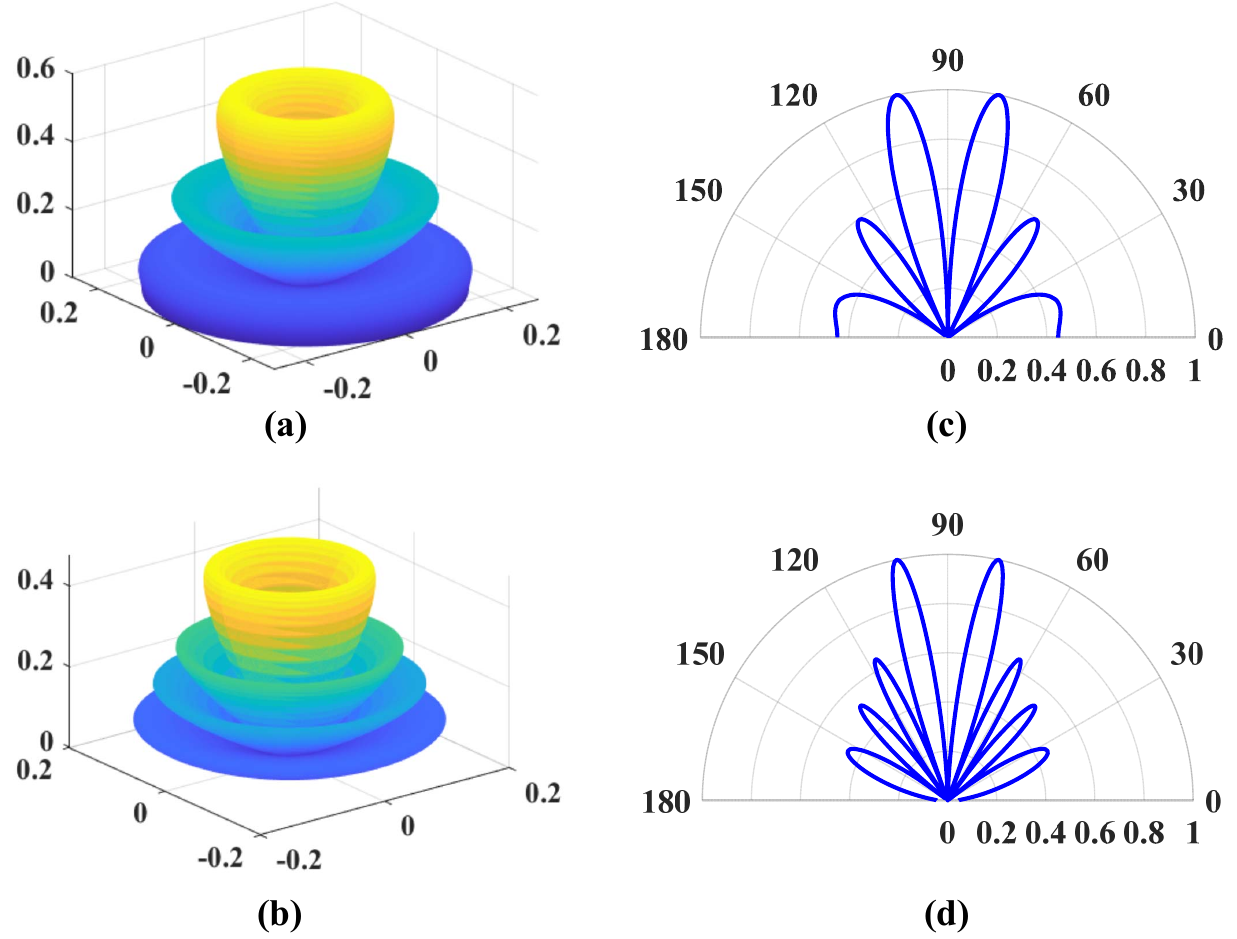}
\caption{The numerical calculation results by MATLAB. (a), (b) 3D radiation patterns of OAM Mode +1 and Mode +2; (c), (d) 2D results of OAM Mode +1 and Mode +2 by Bessel function.}
\label{OAMs_revised}
\end{figure}

\begin{figure}[hbt]
\centering
\includegraphics[width=3.3in]{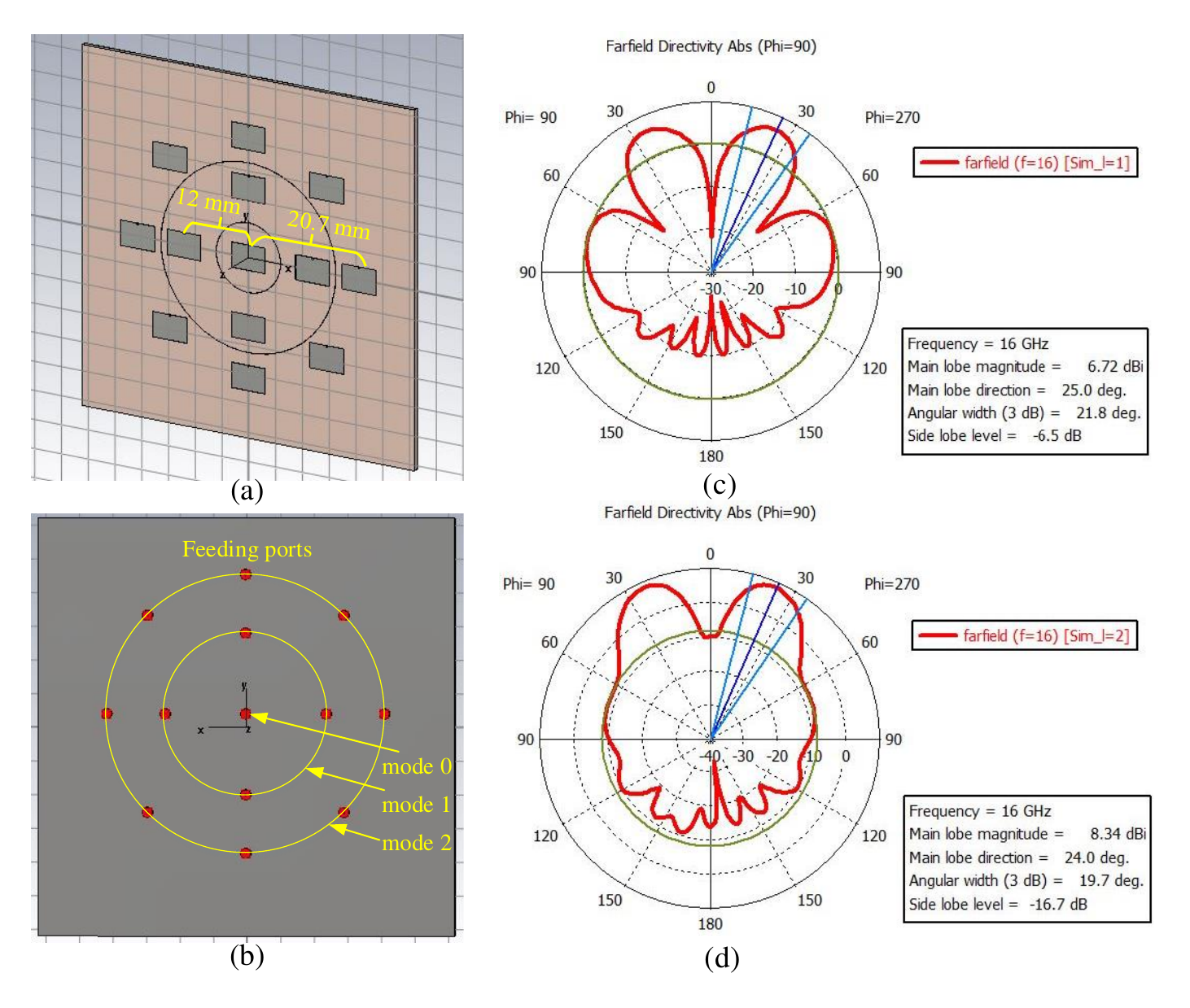}
\caption{Concentric UCA-based OAM generator by CST studio. (a) front side (b) back side, and the CST farfield pattern for (c) Mode +1 (d) Mode +2.}
\label{OAMs_fig4}
\end{figure}

So far, the detailed design process of the multi-mode OAM beam transmitter can be realized step by step as the following.
\begin{itemize}
\item First, for the concentric UCA, the radii of different ring arrays can be calculated by (\ref{eq4}), which gives a guide for the next array arrangement in the full-wave simulation. According to (\ref{eq3}), the Bessel function package that comes with MATLAB can perform numerical simulations for various OAM modes, so as to validate the impact of adjusting the array radii on the beam divergence angles, which is shown in Fig. \ref{OAMs_revised}.
\item Then, the microstrip patch antenna is designed in the CST Studio Suite with the best impedance matching feed position. The same patches are arranged on the circles with different radii according to the calculation results of MATLAB in the previous step. By finely adjusting the radii of different circles, radiation patterns with approximately the same beam divergence angle can be obtained, as shown in Fig. \ref{OAMs_fig4}.
\item Furthermore, the concentric UCA antennas are processed and different phase-shifting networks are fabricated for different OAM modes. Here, for convenience and without loss of generality, we constructed the OAM feed network by means of power dividers with RF delay lines.
\end{itemize}

Both Fig. \ref{OAMs_revised} and Fig. \ref{OAMs_fig4} verify that different OAM modes can share almost the same beam divergence angle. Meanwhile, the radiation patterns obtained through various simulation methods exhibit slight differences. As we know, a typical Bessel function has a definite shape. Therefore, in fact, during the MATLAB simulation, the sidelobes of different OAM patterns are determined by the forms of the $l$-order Bessel functions. These results are obtained assuming that all array elements are omnidirectional radiation and that the number of elements is infinite, which can only be used as a preliminary reference to guide the design of the actual prototype. However, in practice, during the CST full-wave simulation and antenna processing, we can only use a finite number of array elements, and the pattern of each element is also directional. As shown in Fig. \ref{OAMs_fig4}, four rectangular patch elements are used to generate OAM beam mode 1 (inner ring), and 8 patches to generate mode 2 (outer ring). The central patch is for mode 0 (plane wave) radiation. Hence, the entire array pattern is jointly influenced by the array element spacing, the coupling between elements, and even the pattern of each element (if the number of elements in the array is small). In addition, (\ref{eq1}) enlightens us that the phase delay of feeding signals ${{\vec A}_{l,n}} = {I_{l,n}}{e^{j{\alpha _{l,n}}}}$ between different array elements will also affect the sidelobes of the pattern \cite{yufei-EL}.

Actually, besides the concentric UCA generator, multi-mode OAM beams can also be generated at microwaves by using different approaches. In particular, Burghignoli $\textit{et al.}$ \cite{R9} propose to realize the consistent divergence angle for OAM beams of different orders without changing the array radius. They focus on exciting cylindrical leaky waves by feeding a Fabry-Perot cavity with a single circular dipole array, which appears to reduce the number of excitation elements, thereby increasing the efficiency of OAM multi-mode generation. Compared with the concentric UCA, the Fabry-Perot cavity has a more complex structure and does not have the low-profile characteristic of a patch array, and higher requirements for the fabrication and assembly of the actual system. Specifically, in communication application scenarios, the concentric UCA-based OAM generation system is more convenient for researchers to do complex digital-analog hybrid beamforming design. Therefore, this kind of scheme is adopted in many practical system implementations \cite{NTT}, \cite{cheng1}. However, it is undeniable that in order to generate multi-mode OAM beams, whether concentric or single UCA, they both require complex phase-shifting networks customized for the array, e.g., the Butler matrix \cite{Bulter}. Hence, how to apply the RF structures to the generation and reception of actual communication signals is still highly worth more in-depth exploration.

\section{Circular Airy Wavefront Tailored Multi-mode Convergent Transmission Scheme} \label{sect3}
Obviously, it can be seen from (\ref{eq4}) that the higher-order Bessel beams carry the OAM property, which is associated with the inclined wavefronts of such beams \cite{R6}. In fact, the ideal Bessel beam is a kind of non-diffracting wave itself \cite{R4}, or commonly known as localized waves, which are indeed able to resist diffraction for a long distance \cite{R3}. Initially, the researchers found that the energy of this localized wave is confined to the propagation axis, which can counteract the diffraction effect well and benefit remote power transmission or wireless communications \cite{R4}, \cite{R3}. However, an ideal non-diffracting wave only can be generated through an infinite amount of energy and unlimited radiating apertures. In practice, only finite-size finite-energy generators are available for obtaining the pseudo-localized wave, e.g., the quasi-Bessel beam, which has a limited non-diffracting propagation depth \cite{R5}. As we know, the high-order quasi-Bessel beams are always accompanied by OAM, which can maintain a nearly stable EM field for a considerable distance during microwave radio transmission \cite{R6}, \cite{R5}. Hence, many methods have been proposed for the high-order quasi-Bessel beams generation, e.g., reflect arrays \cite{lilong}, metalenses \cite{wuqun}, etc.

Compared with Bessel beams, the Airy beam is also an important member of the non-diffracting wave family \cite{R4}. The Airy beam has the unique feature of abrupt auto-focusing during propagation \cite{R4}, which can bring particular benefits when designing communication scenarios. Specifically, the circular Airy beam is a kind of 2-D Airy beam, which can maintain quite low-intensity profiles during propagation until its energy can be abruptly focused at the focal point \cite{xiuping}, \cite{R10}. Hence, if we combine the quasi-Bessel beams accompanied by OAM and the circular Airy beams together, we can obtain both the OAM and the auto-focusing properties during wireless propagation. It is worth noting that, based on this scheme, the receiver only needs to be placed at the same focal energy convergence region to sample signal information from different OAM modes simultaneously. Similarly, due to the finite generation aperture, an ideal Airy beam does not actually exist \cite{Airy-Nanjing}. To make it experimentally feasible, an exponential decay factor $\alpha$ should be introduced to the Airy beam. To adapt to the cylindrically symmetric morphology of the UCA radiation source, a radially symmetric circular Airy function is used for wavefront modulation of the transmission beam. According to \cite{R10}, \cite{mmWave}, the electric field of circular Airy beam with the finite-energy packet at the starting plane, i.e., ${z {\rm{ = }}0}$, can be expressed as
\begin{equation} \label{eq5}
{\Psi _{{\text{circular}}}}(z{\text{ = }}0,r) = {\text{Ai}}\left[ {\beta \left( {{r_0} - r} \right)} \right]\exp \left[ {\alpha \beta \left( {{r_0} - r} \right)} \right],
\end{equation}
where, $0 < \alpha  < 1$, $z$ denotes distance along the direction of propagation, $r$ is the radial coordinate that can also be rewritten as ${r = \sqrt {{x^2} + {y^2}} }$ in the Cartesian coordinate system, ${\rm{Ai}}\left[  *  \right]$ denotes the Airy function, ${{r_0}}$ is a parameter related to the initial radius parameter of the circular Airy beam, $\beta$ is a constant that denotes the scaling factor of transverse with units of ${{\text{m}}^{ - 1}}$. Specifically, $R \approx \left( {{r_0} + {1 \mathord{\left/ {\vphantom {1 \beta }} \right. \kern-\nulldelimiterspace} \beta }} \right)$, which physically represents the radius of the highest intensity circular Airy beam \cite{Observation}, \cite{mmWave44}.

\begin{figure}[htb]
\centering
\includegraphics[width=3.4in]{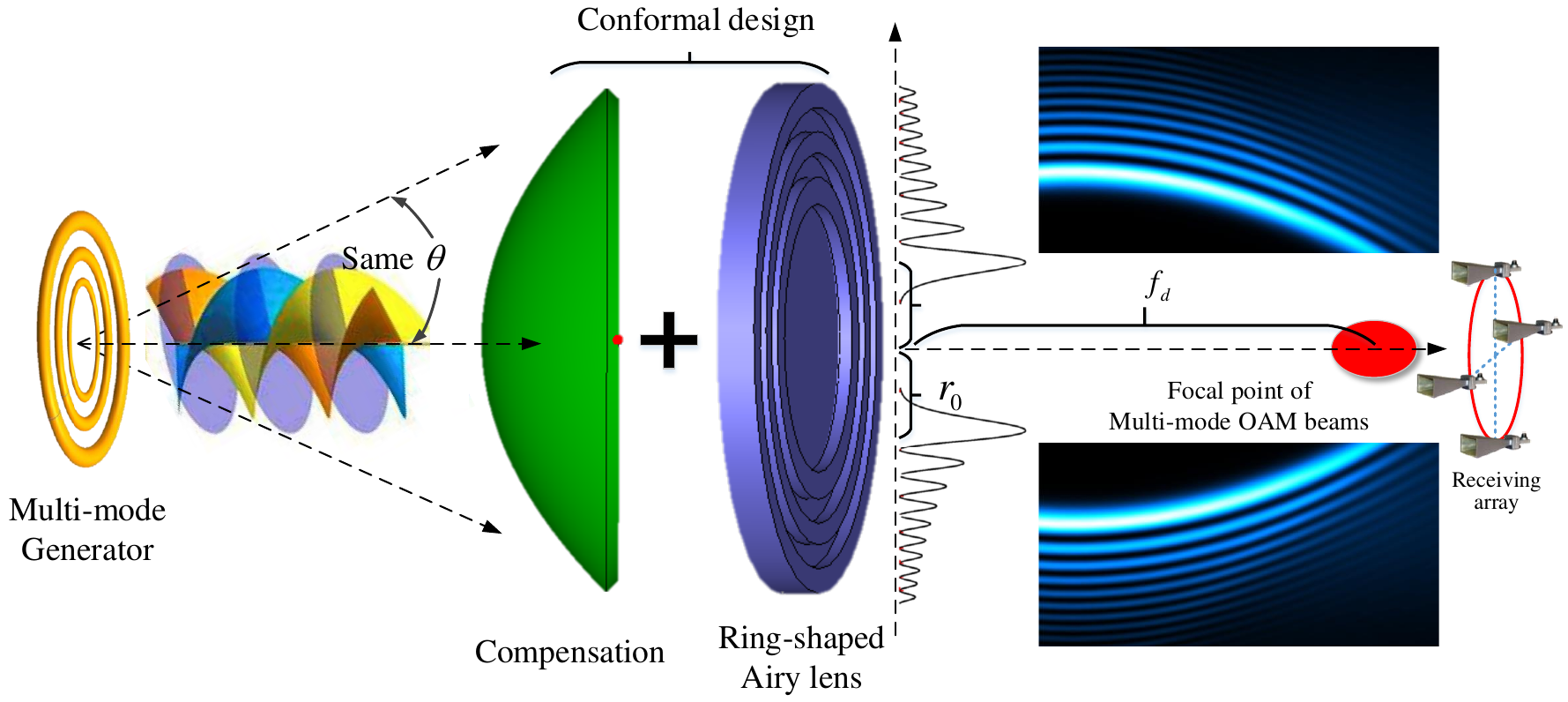}
\caption{OAM beams tailored by ring-shaped Airy wavefront in propagation.}
\label{fig4}
\end{figure}

Previously, researches about Airy beams were primarily focused in the optical field. In recent years, the characteristics of Airy beams in microwave frequencies have gradually attracted the interest of researchers. In \cite{R11}, a microwave 2-D Airy beam is experimentally generated for a frequency range 5.82-5.9 GHz, by employing a single-layer metasurface based on a square C-shaped complementary split-ring resonator. Similarly, researchers from \cite{xiuping} also propose a millimeter-wave Airy beam excitation scheme combined with a single-mode OAM phase distribution by using a single-layer hexagonal lattice metasurface. Furthermore, in addition to passive metasurface, a new method for generating Airy beams based on active microstrip patch antenna array excitation is proposed in \cite{R12}, which further broadens the application scenario of quasi-nondiffractive Airy beams. It is worth noting that, another convenient way to modulate the EM waves is to make them pass through a dielectric material with specific refractive indices. Based on the self-accelerating property of the Airy beam, and considering the inverted cone structure of OAM beams, a ring-shaped Airy lens fabricated by a kind of dielectric material has been proposed in this paper, which is shown in Fig. \ref{fig4}. This lens can realize the wavefront tailoring of multi-mode OAM beams simultaneously, so as to achieve the effect of energy convergence.

In order to generate the Airy wavefront, the phase profile as a function of radial must satisfy ${\varphi _{{\rm{Airy}}}} = \arg \left[ {{\Psi _{{\rm{circular}}}}(\kappa {\rm{ = }}0,r)} \right]$. Moreover, in the actual transmission system, the wavefront steering device is typically located near the OAM antennas. That is, due to the different propagation paths, the phases of EM waves arriving at different radii of the ring-shaped Airy lens are also inconsistent. Hence, in addition to the Airy beam phase modulation, additional compensation for the phase distribution on the lens is also required. Assuming that the vertical distance between the OAM antenna and the EM lens is ${d_0}$, the center of the OAM generator is taken as the origin, $x = {d_0}\sin \theta \cos \phi $ and $y = {d_0}\sin \theta \sin \phi $ denote the coordinates of any position on the plane where the lens is located in the Cartesian coordinate system. As a result, the total phase shift ${\varphi _{{\rm{total}}}}$ of the lens after compensating can be denoted as
\begin{equation} \label{eq6}
{\varphi _{{\rm{total}}}} = {\varphi _{{\rm{Airy}}}} + k\left( {\sqrt {d_0^2 + {x^2} + {y^2}}  - {d_0}} \right),
\end{equation}
After the wavefront steering, the electric field vector at any observation position in free space can be denoted as
\begin{equation}
\begin{gathered}
  {{\vec E}_p} = {{\vec A}_l}{{\vec E}_l}\left( {d,\theta ,\phi } \right){e^{j{\varphi _{{\text{total}}}}}} \hfill \\
  {\kern 1pt} {\kern 1pt} {\kern 1pt} {\kern 1pt} {\kern 1pt} {\kern 1pt} {\kern 1pt} {\kern 1pt} {\kern 1pt} {\kern 1pt} {\kern 1pt} {\kern 1pt}  = {{\vec A}_l}\gamma {J_l}\left( {k{R_l}\sin \theta } \right){e^{j\left( {l\phi  + {\varphi _{{\text{total}}}}} \right)}}, \hfill \\
\end{gathered}
\end{equation}
where ${{\vec A}_l}$ is the vector signal modulated by each OAM channel. As we know, Airy beams are self-accelerating. Specifically, for the ring-Airy beam, the radius of the ring formed by the main lobe in space will suddenly decrease after a certain propagation distance, and the energy will converge and increase. The propagation distance from the Airy lens to the converging focal point is recorded as the focal length ${f_d}$. At the focal point, the maximum deflection of the main lobe of the Airy beam is denoted as ${x_d}$. According to \cite{mmWave}, ${x_d}$ can be calculated as
\begin{equation}
{x_d} = \frac{{{\beta ^3}f_d^2}}{{4{k^2}}}.
\end{equation}
On the other hand, the initial radius of the whole ring-Airy beam is approximately ${r_0} + {1 \mathord{\left/
{\vphantom {1 \beta }} \right.
\kern-\nulldelimiterspace} \beta }$ \cite{mmWave44}. In principle, the radius of the ring-Airy beam shrinks from ${r_0} + {1 \mathord{\left/
{\vphantom {1 \beta }} \right.
\kern-\nulldelimiterspace} \beta }$ to $0$, which is equivalent to the deflection of ${x_d} = {r_0} + {1 \mathord{\left/
{\vphantom {1 \beta }} \right.
\kern-\nulldelimiterspace} \beta }$, then the maximum convergence propagation distance ${f_d}$ can be calculated as \cite{mmWave}
\begin{equation}
{f_d} = \sqrt {4{k^2}\left( {{r_0} + {1 \mathord{\left/
 {\vphantom {1 \beta }} \right.
 \kern-\nulldelimiterspace} \beta }} \right)/{\beta ^3}}.
\end{equation}
Hence, ${f_d}$ can be flexibly adjusted by picking these appropriate parameters $\beta $, ${{r_0}}$, and $\lambda $.

\section{Prototype and Performance Evaluation} \label{sect4}
\begin{figure}[htb]
\centering
\includegraphics[width=3.4in]{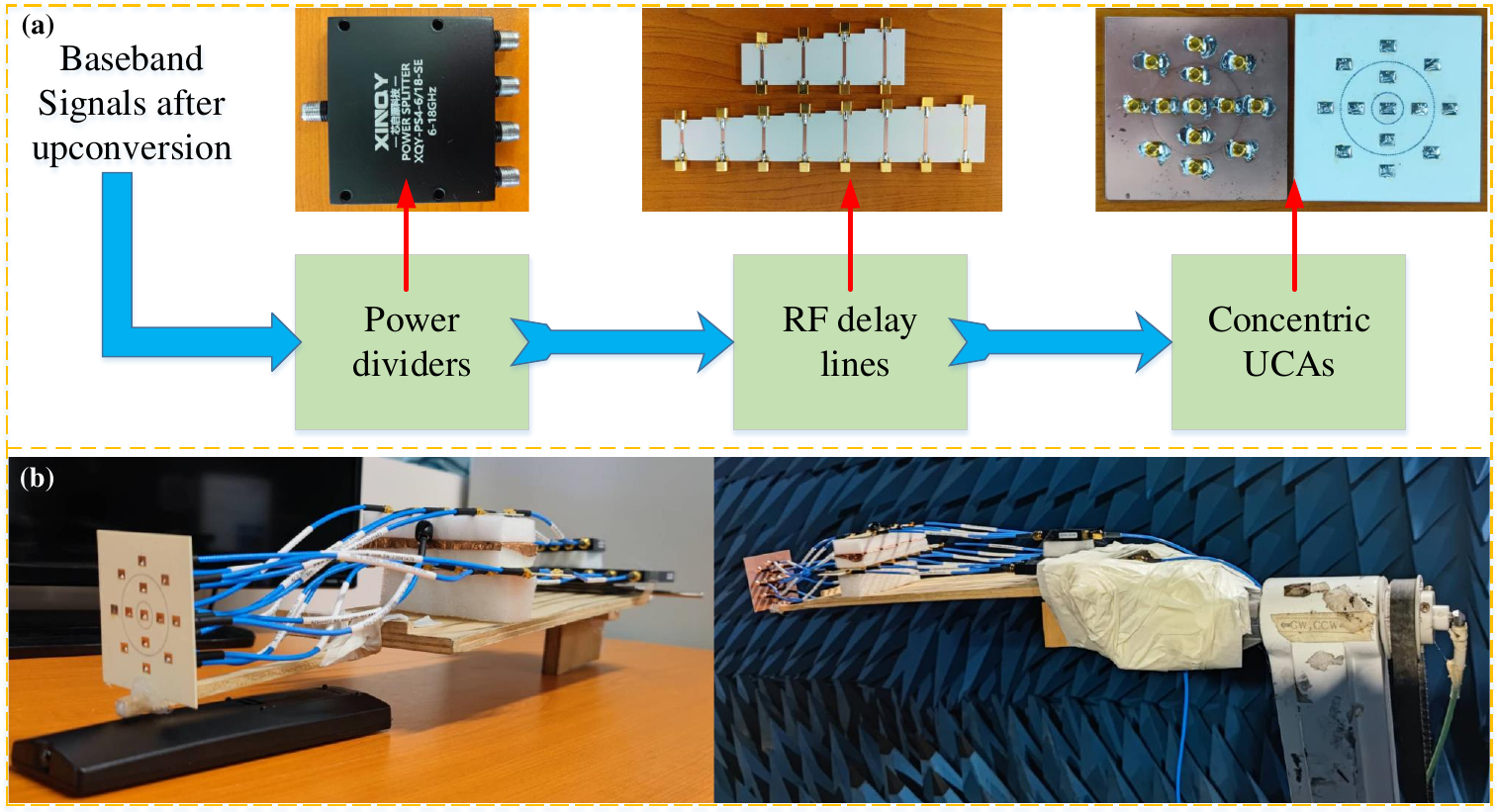}
\caption{Multi-mode OAM RF transmitter (a) configuration, (b) fabrication and experiment scenario.}
\label{new_fig6}
\end{figure}

At the transmitter, the patch elements in each UCA ring are fed with the same signal, but with a successive delay from element to element. The beamforming process for the concentric UCAs to generate multi-mode OAM beams can be performed by a discrete Fourier transform (DFT) \cite{reviewer2}. The transmission vector can be expressed as
\begin{equation} \label{neweq1}
{\mathbf{s}} = {{\mathbf{W}}_N}{\mathbf{Px}},
\end{equation}
where, ${\mathbf{x}}$ is the data symbols mapped to each OAM mode, ${\mathbf{P}}$ denotes the power allocation matrix, which ensures that the received power is almost the same for all transmitted OAM modes, ${\mathbf{W}}_N$ is the DFT matrix, $N$ is the element number in each UCA ring. Assuming that each circle of UCA has the same number of elements $N$ in (\ref{neweq1}), then the phase distribution corresponding to the $l$-order OAM mode can be expressed as
\begin{subequations} \label{neweq2}
\begin{align}
{{\mathbf{w}}_n} = \frac{1}{{\sqrt N }}{\left[ {1,{w_n},w_n^2, \ldots ,w_n^{N - 1}} \right]^T}, \\
{w_n} = \exp \left( { - j2\pi {{\left( {n - 1} \right)} \mathord{\left/
 {\vphantom {{\left( {n - 1} \right)} N}} \right.
 \kern-\nulldelimiterspace} N}} \right),
\end{align}
\end{subequations}
where, $n \in \left\{ {1,2, \ldots ,\left\lfloor {{N \mathord{\left/ {\vphantom {N 2}} \right. \kern-\nulldelimiterspace} 2}} \right\rfloor } \right\}$. In practice, there are various methods to realize this kind of successive phase delay, and the Butler matrix is a well-established and efficient integrated approach that has been validated in \cite{Bulter}. After integrating the design complexity and processing cost factors, a more concise and economical approach has been adopted in our experiments. As shown in Fig. \ref{new_fig6}(a), after up-conversion, baseband signals are connected to the RF delay lines modules via different power dividers. Depending on the lengths varying, different modules can generate various phase delays for each OAM mode (e.g., $2\pi$ for Mode +1, $4\pi$ for Mode +2, etc.), which are then connected to the UCA's SMP connectors. Fig. \ref{new_fig6}(b) shows the fabrication and experiment scenario in the microwave anechoic chamber.

In addition to the multi-mode transmitter, the generation and control of Airy beams also play a pivotal role in our design. In theory, the generation of an Airy beam involves both amplitude and phase modulation. However, previous studies have discovered that Airy beams can be approximated quite well by phase-only modulation \cite{phase-only1}, \cite{phase-only2}. Since the amplitude modulation suffers from low efficiency and makes the system setup complex. Hence, to reduce the complexity of the ring-shaped Airy lens, the phase-only modulation is utilized in our design. As we know, complex lens shapes can be realized in a flexible and timely manner using Computer Numerical Control (CNC) machines or 3D printing techniques. To quickly verify the correctness of the design method discussed in Sect. \ref{sect3}, in this paper, we first use a High-Density Polyethylene (HDPE) dielectric material to achieve wavefront modulation of multi-mode OAM beams.

\begin{figure}[htb]
\centering
\includegraphics[width=3.4in]{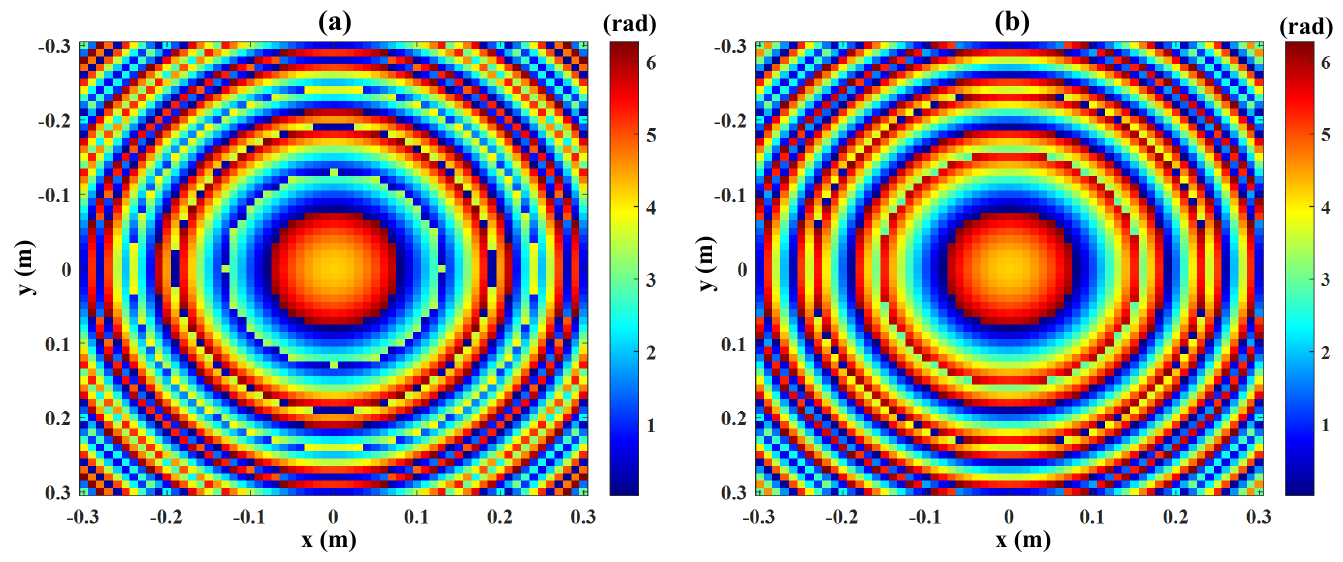}
\caption{Phase distributes ($0 \sim 2\pi$ radians) of the Airy EM lens via different focus ranges. (a) ${f_d} = 1.17$ m; (b) ${f_d} = 2.09$ m.}
\label{phase}
\end{figure}

Different heights of HDPE materials with uniform density will produce different phase delay variations for EM waves during propagation. Upon transmission through the phase plate, a beam of wavelength $\lambda$ is subject to a phase delay, and the relationship between step thickness and the phase shift can be deduced as \cite{SPP}
\begin{equation}
\Delta h = \left( {\frac{{{\varphi _{{\rm{total}}}} + {\phi _0}}}{{2\pi }}} \right) \times \frac{\lambda }{{\eta  - 1}},
\end{equation}
where, ${{\phi _0}}$ is an arbitrary phase constant, $\eta  = \sqrt {{\varepsilon _r}{\omega _r}} $ denotes the refractive index, ${\varepsilon _r}$ is the relative dielectric constant, ${\omega _r}$ is the relative magnetic permeability, for HDPE, ${\varepsilon_r}  \approx 2.9$, ${\omega_r}  \approx 1$. The phase distributes of the Airy EM lens after compensating for different focus ranges are shown in Fig. \ref{phase}.

\begin{figure}[htb]
\centering
\includegraphics[width=3.4in]{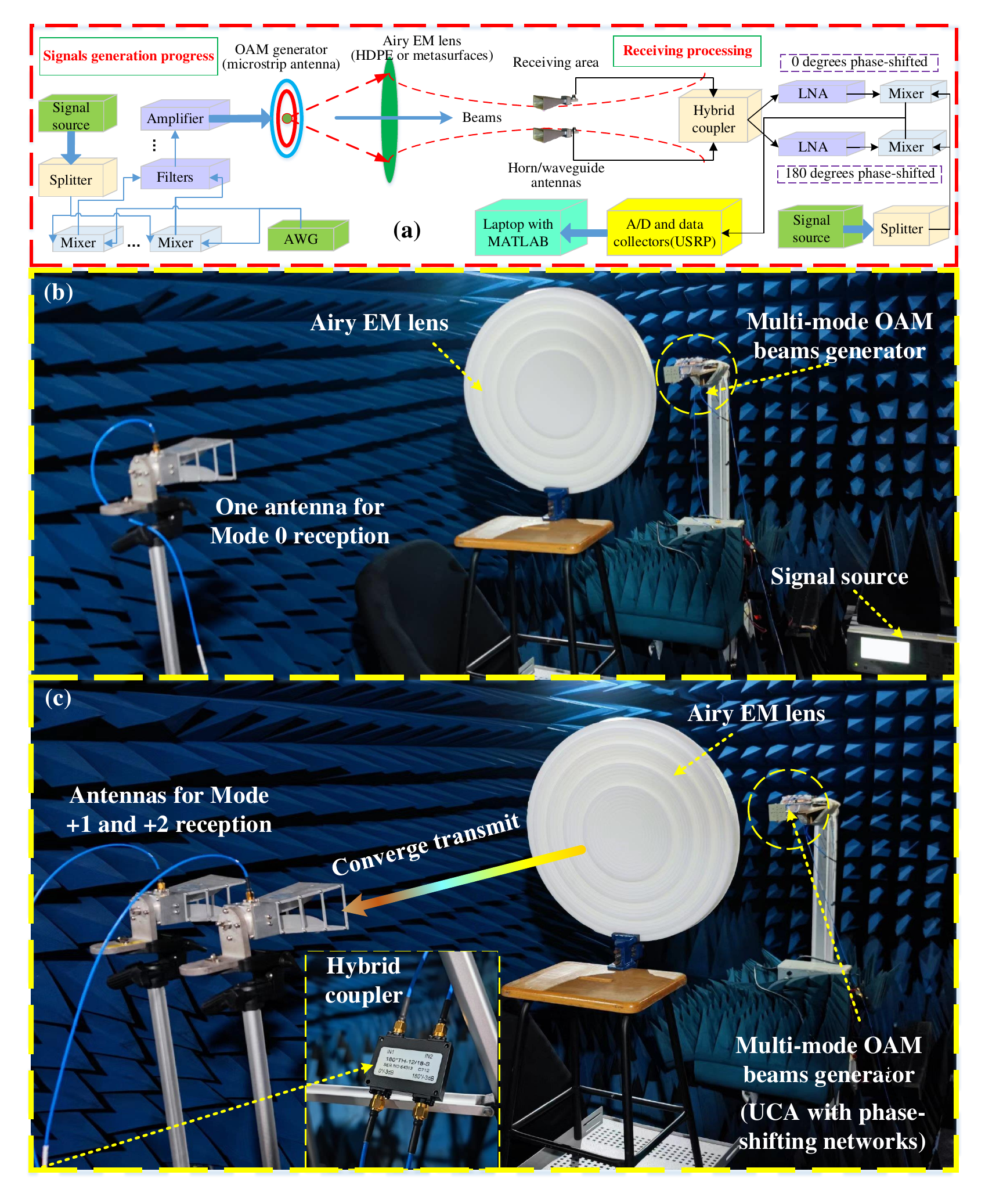}
\caption{Experiment prototype of multi-mode OAM beams convergent transmission and receiving. (a) Schematic diagram; (b) Measurement scene of Mode 0; (c) Measurement scene of Mode +1 and +2}
\label{prototype}
\end{figure}

The transmission experimental prototype is shown in Fig. \ref{prototype}. The transmitter consists of multiple UCA antennas, which are connected to independent frequency conversion links and baseband signal processors. In order to reduce the complexity of the experiment, the transmitter adopts the most mature UCA forms of annular nesting, and realizes vortex beamforming via phase-shifting networks, which can generate 5 different modes of OAM beams, i.e., $\left\{ {{\rm{ - }}1,{\rm{ - }}2,0,{\rm{ + }}1,{\rm{ + }}2} \right\}$. The Arbitrary Waveform Generator (AWG, Tectronix AWG70001A) is used to generate modulated Intermediate Frequency (IF) signal, which is up-converted, amplified, filtered, and then fed to the RF end. In the experiment, according to (\ref{eq6}), the phase compensation of the wavefront steering lens is designed according to its relative spatial position to the center of the OAM generator, while superimposing the phase distribution properties of the Airy beams. As shown in Fig. \ref{prototype}, the transmission experiment was carried out in a microwave anechoic chamber at the certain frequency of 16.1 GHz. Considering the size of the anechoic chamber, the distance ${d_0}$ between the UCAs generator and the EM lens is set to 0.9 m, and ${r_0} = 0.04$, $\beta  = 29$, the maximum thickness of the EM lens is 0.032 m.

\begin{figure*}[htb]
\centering
\includegraphics[width=6.6in]{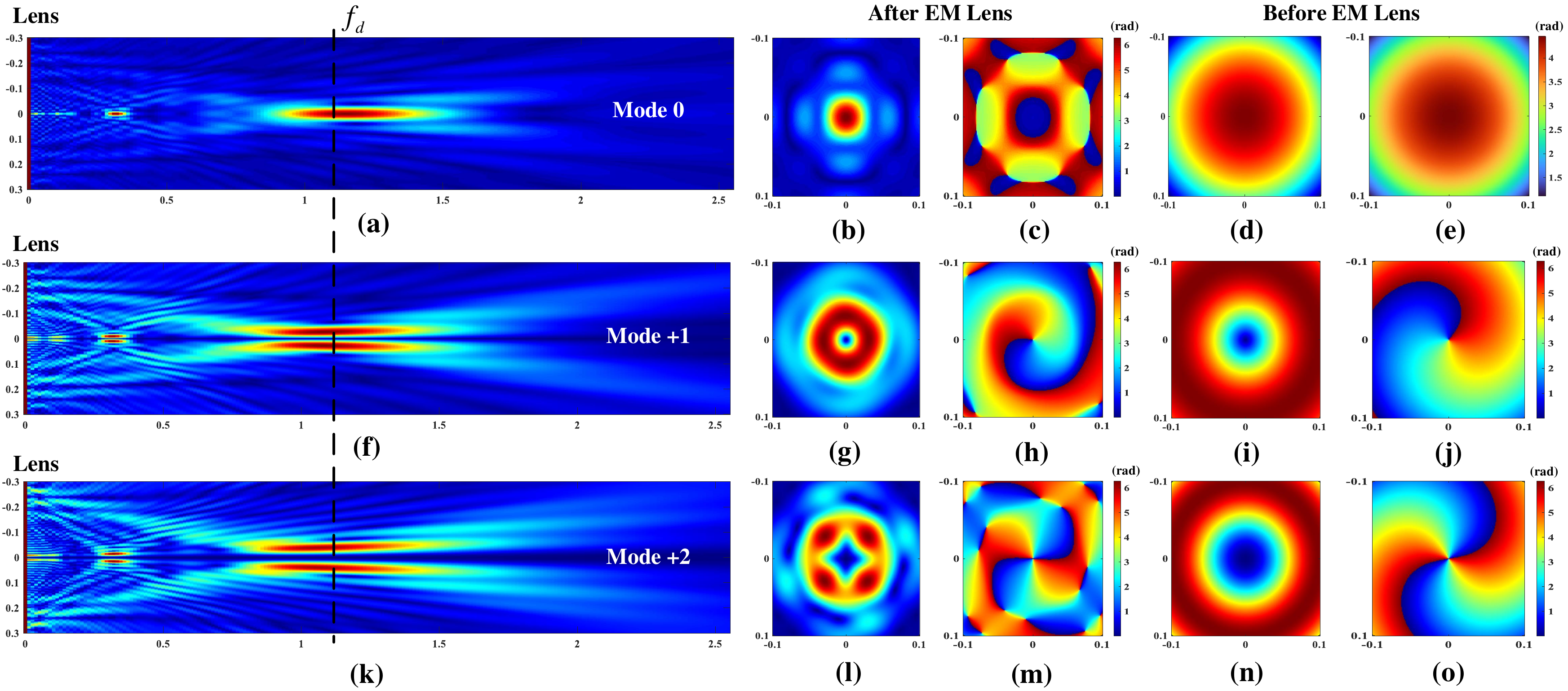}
\caption{Simulation results of the transmission experiment. Energy distributions of (a) Mode 0,(f) Mode +1,and (k) Mode +2 in the longitudinal plane along the propagation direction after the Airy EM lens. (b) Amplitude and (c) phase of Mode 0 in the transversal plane at focal point ${f_d}$ after EM lens. (d) Amplitude and (e) phase of Mode 0 in the transversal plane at focal point without EM lens. (g) Amplitude and (h) phase of Mode +1 in the transversal plane at focal point after EM lens. (i) Amplitude and (j) phase of Mode +1 in the transversal plane at focal point without EM lens. (l) Amplitude and (m) phase of Mode +2 in the transversal plane at focal point after EM lens. (g) Amplitude and (h) phase of Mode +2 in the transversal plane at focal point without EM lens.}
\label{OAM-Airy}
\end{figure*}

Firstly, the numerical calculation results are shown in Fig. \ref{OAM-Airy}. The OAM beams of different modes are tailored by the Airy wavefront and converge to a position near the propagation axis at ${f_d} = 1.17$ m after the Airy EM lens, i.e., the distance between the sampling receiving area and the OAM transmitter is about ${d_0} + {f_d} = 2.07$ m. As shown in Fig. \ref{OAM-Airy}, the varieties of the multi-mode OAM beams with or without wavefront steering through the Airy EM lens during propagation are compared at the same receiving area. It can be seen that after the phase compensation and convergence of the Airy EM lens, the OAM beam can be converged to a specific spatial position along the propagation axis, the energy is more concentrated, and the phase distribution characteristics of the spiral wavefront can still be maintained. For the receiving end, this transmission scheme not only enables the receiving antennas to simultaneously receive multiple OAM modes in a same limited space area, but also effectively overcomes the problem of OAM beam energy divergence, greatly improving the pre-detection SNR.

Then, the actual receiving power variation and isolation between multiplexed OAM channels are measured. During the experiment, the generator activates each OAM mode with the same transmitting power. As shown in Fig. \ref{prototype}(a)(c), the receiving end uses two horn antennas, which are placed at the focal point (according to the numerical results in Fig. \ref{OAM-Airy}), and connected to the RF receiver after passing through a ${180^ \circ }$ hybrid coupler. According to the differences in the spatial phase distribution of OAM modes 1 and 2, we only need to place two horns at both ends of the diameter of the annular phase plane. As shown in Fig. \ref{prototype}(c), near the focal point of the lens, two points with the strongest received power and the farthest distance from each other can be found through measurement, which are the locations of the receiving antennas. The distances from the two antennas to the lens are equal, i.e., 1.17 m, and these two antennas are approximately 0.26 m apart, which guarantees that two signals with different vorticity can be sampled. The sampling signals pass through the hybrid coupler synchronously, for Mode 1, the two signals are coherently superimposed after being phase-shifted by 180 degrees; while for Mode +2, in principle, the two sampling signals are directly coherently superimposed with the same phase (0 degrees phase-shifted). In this way, analog RF separation detection of different OAM modes can be implemented with very low computational complexity. The actual experiment scenario is shown in Fig. \ref{prototype}(c), and the measured results are illustrated in Table \ref{tab1}.
\begin{table*}[htbp]
\caption{Isolation measurement results of 3 different order OAM modes at the same longitudinal reception distance}
\begin{center}
\begin{tabular}{c|c|c|c|c|c|c|c|c}
\toprule
\textbf{Receiving} & \multicolumn{3}{c|}{\textbf{Transmitting modes with lens}} & \multicolumn{3}{c|}{\textbf{Transmitting modes without lens}} & \multicolumn{2}{c}{\textbf{Receiving isolations (with lens)}} \\
\cmidrule{2-7}
\textbf{modes power} & ${l_0} = 0$ & ${l_1} = +1$ & ${l_2} = +2$ & ${l_0} = 0$ & ${l_1} = +1$ & ${l_2} = +2$ & \multicolumn{2}{c}{\textbf{to other different Tx modes}} \\
\midrule
${l_0} = 0$ & -22.3 dBm & -37.9 dBm & -37.4 dBm & -35.8 dBm & -53.2 dBm & -52.7 dBm & 15.6 dB ($l_1$) & 15.1 dB ($l_2$)\\
\midrule
${l_1} = +1$ & -39.2 dBm & -24.8 dBm & -36.7 dBm & -51.6 dBm & -38.6 dBm & -46.6 dBm & 14.4 dB ($l_0$) & 11.9 dB ($l_2$) \\
\midrule
${l_2} = +2$ & -37.6 dBm & -35.8 dBm & -24.5 dBm & -50.2 dBm & -45.3 dBm & -37.4 dBm & 13.1 dB ($l_0$) & 11.3 dB ($l_1$) \\
\bottomrule
\end{tabular}
\label{tab1}
\end{center}
\end{table*}

To clarify, let's take the second row in Table \ref{tab1} as an example. The receiving horns with the ${180^ \circ }$ hybrid coupler are preset to receive the signal of mode ${l_1}$. When other OAM modes are transmitted, e.g., the receiver performs phase shifting for modes ${l_0} = 0$, ${l_2} = +2$ according to the phase distribution of ${l_1} = +1$ (i.e., ${180^ \circ }$ phase-shifting and combination), and the receiving energy incoherent decrease to -39.2 dBm, -36.7 dBm respectively under the condition with the lens. In other words, only when ${l_1} = +1$ mode is transmitted, it matches the ${180^ \circ }$ phase-shifting operation and the energy coherent creases to -24.8 dBm. As for the channel isolations, Mode $0$ performs the best, which is thanks to the fact that the receiving antenna of Mode $0$ is located at the midpoint and is spatially isolated from the radiation direction of the main lobes of the other modes. The isolation between Modes $+1$, $+2$ is over 11 dB, which would be mainly impacted by the imperfect fabrications (e.g., generation antennas, Airy lens, hybrid coupler, etc.), the misalignment between the Tx/Rx antennas, and the inconsistency of link insert-loss.
To a certain extent, the measurement results verified that the isolations between OAM channels can be implemented by pure analog hardware operation, which greatly reduces the computational complexity of the receiver. Moreover, comparing the results with or without the Airy EM lens, it is obvious that after the convergence of the lens, the energy of the received signals is increased by at least 12 dB for all kinds of modes within the same receiving area.

\begin{figure}[!t]
\centering
\includegraphics[width=3.2in]{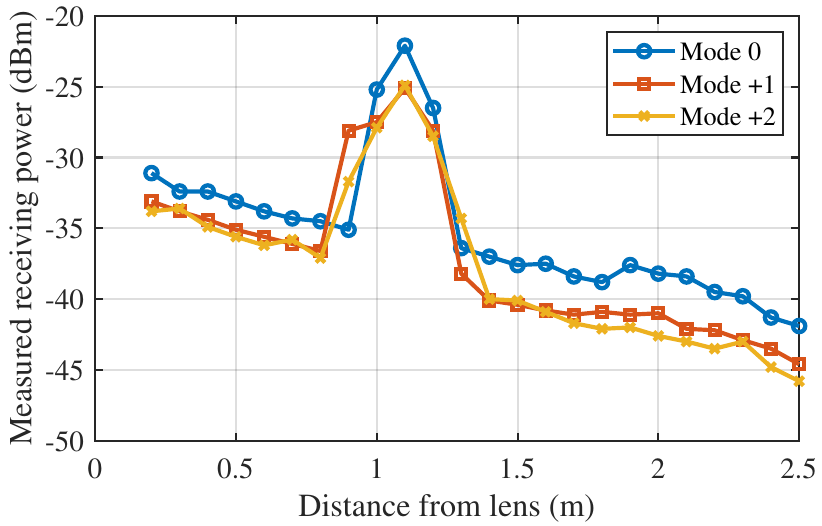}
\caption{Measurement results of receiving power vs. propagation axis.}
\label{power}
\end{figure}

Figure \ref{power} shows the measured receiving power along the propagation direction. Different modes are enabled one by one during the measurement progress. As shown in Fig. \ref{prototype}(b), due to the distinctive central power null of other nonzero OAM modes, the receiving antenna of Mode 0 can be placed at the center of the beams, which will not be affected by other nonzero OAM modes \cite{xuefeng}. The distance from the receiving antenna to the lens is gradually changed to measure the power variation of the beam along the propagation path, while keeping the receiving antenna on the same level as the center of the lens. Then, replace the central antenna with two identical horn antennas, repeat the above process, and obtain the power measurement results of Modes +1 and +2 respectively. Similarly, the height of the two antennas remains the same as the center of the lens, and the antenna spacing remains unchanged at about 0.26 meters. The measurement results are recorded in Fig. \ref{power}. Obviously, as the transmission distance increases, the received power of each mode increases first and then decreases. There is obvious energy convergence at about 1.1 meters from the lens, and the received power is significantly increased, which is consistent with the simulation results in Fig. \ref{OAM-Airy}.

\begin{figure}[!t]
\centering
\includegraphics[width=3.5in]{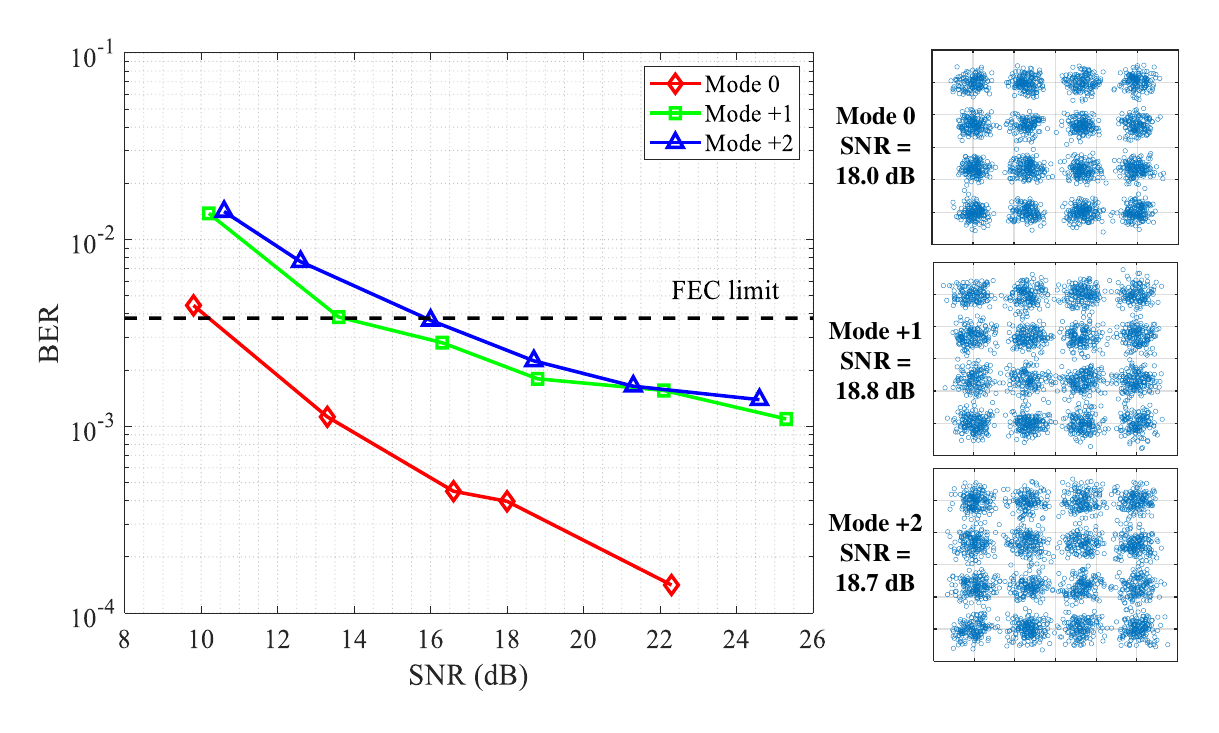}
\caption{The measured BER performances of the 16-QAM multiplexing with Airy lens (without channel coding and equalization).}
\label{BER}
\end{figure}

Furthermore, after the power loss and isolation measurement, the data multiplexing experiment is also implemented through these 3 different modes. Specifically, at the transmitter side, each mode sends a pre-defined random sequence with 16-QAM modulation repeatedly, which makes the demodulation progress much easier. In other words, the receiver only needs to extract one complete sequence from the whole receiving data and compare it with the original transmitting sequence. The whole demodulation process is to first store the received data and then perform off-line calculation, which has high stability and repeatability. The basic experimental setup is shown in Table \ref{tab2}.
\begin{table}[htb]
\caption{Some main experiment parameters.}
\begin{center}
\begin{tabular}{c|c|c}
\toprule
\textbf{Parameter} & \textbf{Value} & \textbf{Dimension} \\
 \midrule
 Central carrier frequency & 16.1 & GHz \\
 Intermediate frequency & 430.0 & MHz \\
 Baseband data bandwidth & 1.0 & MHz \\
 Amplifier noise figure & 6.5 & dB \\
 Amplifier gain & 26.1 & dB \\
 Airy lens diameter & 0.6 & m \\
 Multiplexing OAM modes & 0,+1,+2 & - \\
 Baseband modulation & 16-QAM & - \\
 \bottomrule
\end{tabular}
\end{center}
\label{tab2}
\end{table}
A single antenna is put in the midpoint between the two horn antennas to receive Mode 0 signal, and the other two antennas are connected to the ${180^\circ }$ hybrid coupler to separate signals from Modes $+1$, $+2$. It is worth noting that during the data multiplexing experiment, three different OAM modes are activated simultaneously for the purpose of observing the demodulation performance in the presence of inter-mode interference. The Signal to Interference and Noise Ratio (SINR) can be denoted as
\begin{equation}
{\eta _{{\text{SINR}}}} = \frac{{{S_{{\text{each mode}}}}}}{{{N_{{\text{noise}}}} + {I_{{\text{inter - mode interference}}}}}},
\end{equation}
where, ${{S_{{\text{each mode}}}}}$ denotes the useful signal power carried by each mode, which can be controlled by adjusting the amplitude of the IF signal at the output of AWG, ${{N_{{\text{noise}}}}}$ equals to $ - 174\left( {{\text{dBm}}} \right) + {\delta _{{\text{NF}}}} + 10{\log _{10}}\left( B \right)$, $B$ is the bandwidth, ${\delta _{{\text{NF}}}}$ is the noise figure of the active component, ${{I_{{\text{inter - mode interference}}}}}$ is the crosstalk between modes which can be calculated from the isolations in Table \ref{tab1}. The variation of BER with SNR can be measured by varying the transmitting power of the useful signals, i.e., the IF signals, during the communication experiments. The results are shown in Fig. \ref{BER}.

The curve of Mode 0 has the best BER performance, thanks to the natural spatial isolation of its beam pattern from the other OAM modes \cite{xuefeng}, which is consistent with the measurement results in Table \ref{tab1}. Moreover, it is obvious that as the SNR increases, the BER curves corresponding to all modes show varying levels of BER floors, which means the mode crosstalk is the main factor limiting the demodulation performance. Considering imperfect practical factors, such as, the fabrication errors of the lens, misalignment between the Tx/Rx antennas, imperfect OAM beams generators, the transmission scheme itself should have a better performance. It is clear that both curves are capable of reaching the Forward Error Correction (FEC) limit ($3.8 \times {10^{ - 3}}$) \cite{FEC}, demonstrating the viability of such a communication link. On the other hand, the mode purity is also an important factor affecting crosstalk. There are many methods to improve the purity of multi-mode generation \cite{NTT,Liang,weite}, which will also be our future work. Moreover, since there is a symmetric spatial structure between the positive and negative modes, for simplicity, only 3 modes $0$, $+1$ and $+2$ are tested in the experiment, and the results of $-1$ and $-2$ will be similar.
The comparison between the proposed scheme and other designs is listed in Table \ref{table3}. To sum up, our proposed scheme enables the dynamic adaptation of different OAM modes and can support simultaneous separation of different modes at the same spatial location, which is more suitable for practical wireless communication architectures.
\begin{table}[htb]
\caption{Comparison between similar designs in the literatures.}
\begin{center}
\begin{tabular}{c|c|c|c}
\toprule
\textbf{Ref.} & \textbf{Multiplexing} & \textbf{Tunable} & \textbf{Demodulation}\\
 \midrule
 \cite{xiuping} & Not support & Mode fixed & NA. \\
 \cite{Airy-Nanjing} & Not support & Mode fixed & NA. \\
 \cite{mmWave} & Not support & Mode fixed & NA. \\
 Our scheme & Achieved & Mode tunable & Simultaneously \\
 \bottomrule
\end{tabular}
\end{center}
\label{table3}
\end{table}

\section{Conclusion} \label{sect5}
A multi-mode OAM beams convergence transmission and co-scale reception scheme is proposed and verified by both simulation and actual experiments. Based on this design, OAM beams of different modes can be transmitted with the same beam divergence angle, and can converge to the same spatial position after being tailored by the ring-shaped Airy wavefront. This improves not only the receiving SNR, but also allows for the simultaneous reception and demodulation of different OAM channels (modes) multiplexed in a limited space area. The measurement isolations between different OAM channels are over 11 dB, which ensures a reliable 16-QAM multiplexing wireless transmission system with acceptable BER results. Compared with other conformal design solutions, the design proposed in this paper separates the multi-mode OAM generation device and the converging lens, which is more suitable for the needs of practical communication scenarios. Based on the reconfigurable antenna technology and Airy functions characteristics, in the future, the OAM modes will also be able to be flexibly adjusted according to the crosstalk, and adapt the Airy lens phase distribution based on changes of the focal point. Meanwhile, the metasurface is being considered to flexibly control various parameters of the Airy EM lens, which will soon be reflected in our future works.


%

%


\ifCLASSOPTIONcaptionsoff
  \newpage
\fi

\end{document}